# System level mechanisms of adaptation, learning, memory formation and evolvability: the role of chaperone and other networks

Dávid M. Gyurkó[1], Csaba Sőti[1], Attila Steták[2] and Peter Csermely[1,1]

[1]Department of Medical Chemistry, Semmelweis University, Budapest, Hungary; [2]Division of Molecular Neurosciences, Faculty of Psychology and Biozentrum, Life Sciences Training Facility, University of Basel, Basel, Switzerland

**Abstract:** During the last decade, network approaches became a powerful tool to describe protein structure and dynamics. Here, we describe first the protein structure networks of molecular chaperones, then characterize chaperone containing sub-networks of interactomes called as chaperone-networks or chaperomes. We review the role of molecular chaperones in short-term adaptation of cellular networks in response to stress, and in long-term adaptation discussing their putative functions in the regulation of evolvability. We provide a general overview of possible network mechanisms of adaptation, learning and memory formation. We propose that changes of network rigidity play a key role in learning and memory formation processes. Flexible network topology provides 'learning-competent' state. Here, networks may have much less modular boundaries than locally rigid, highly modular networks, where the learnt information has already been consolidated in a memory formation process. Since modular boundaries are efficient filters of information, in the 'learning-competent' state information filtering may be much smaller, than after memory formation. This mechanism restricts high information transfer to the 'learning competent' state. After memory formation, modular boundary-induced segregation and information filtering protect the stored information. The flexible networks of young organisms are generally in a 'learning competent' state. On the contrary, locally rigid networks of old organisms have lost their 'learning competent' state, but store and protect their learnt information efficiently. We anticipate that the above mechanism may operate at the level of both protein-protein interaction and neuronal networks.

**Running title:** Chaperone networks in adaptation

[1] Address correspondence to this author at the Department of Medical Chemistry, Semmelweis University, H-1444 Budapest, P. O. Box 260, Hungary; Tel: +36-1-459-1500; Fax: +36-1-266-3802; E-mail: csermely.peter@med.semmelweis-univ.hu

# 1. INTRODUCTION: THE NETWORK APPROACH

We are in the age, when peta-bytes (a million times billion bytes) of biological data become available on the internet. This daunting amount of valuable data requires new methods to explore, understand and utilize. The network approach efficiently selects important from non-important, and defines functionally relevant sets of data in a multi-level hierarchy. Therefore, the network approach emerged as one of the modern, powerful tools to assess key actors and major mechanisms of the regulation and changes of biological functions. When talking about cellular networks, we simplify the multiple dimensions of cellular datasets. In the use of the network approach first we define nodes of the networks (see **Table 1** for a glossary of network-related terms), which may be amino acids, proteins or other macromolecules, cytoskeletal fibers, chromatin segments, cellular organelles, cells or single organisms. As a second step, we define the network edges between the nodes, which are often their physical or functional interactions. Last, we often need to define a weight of the network edge meaning the intensity (strength, probability) of the particular interaction between the two network nodes, or a direction specifying which node is acting on the other *via* the network edge connecting them [1-4].

Acknowledging the enormous progress of the last decade in exploration of biological networks, we must admit that the potential of the network approach is far from being fully utilized. In reality, networks are seldom directed in an unequivocal way. (When children and their parents are talking to each other, it is not always the case that parents influence their children, and the children are not influencing their parents at all.) Moreover, current biological network approaches seldom use signed networks or colored graphs, i.e. an interaction set, where the various types of interactions (e.g. activation or inhibition in case of signed networks and multiple attributes in case of colored graphs) are discriminated. Additionally, conditional edges (meaning edges, which are active only, if one of their nodes accommodated another edge previously) are also very seldom used in current biological network science. Lastly, in many systems sets of nodes are often just simply 'together', making all possible connections with each other. These hypergraphs (where nodes belonging to a set are not linked individually, but are taken together as a group) is also rather seldom used in current biological network science.

We have to warn, that in most biological systems data coverage often has technical limitations, and experimental errors are rather prevalent. Therefore not all of the possible interactions are detected, and a large number of false-positives also appear [4-6]. However, it is often a question of personal judgment, whether the investigator takes only 'high-fidelity' interactions into account, and neglects all others as potential artifacts, or uses the whole spectrum of data considering the weak interactions as low affinity interactions – or as artifacts [3, 7]. Ambiguity tolerance, a major need of scientific endeavor, i.e., a resistance against the extreme simplification of the totality of nature to yes/no answers, becomes especially important, if we work with biological network data.

**Table 1.   A glossary of network-related terms.**

| Network-related term | Explanation |
|---|---|
| node | A network node is a single building block of a network. A node is also called as a network element or vertex. |
| edge | A network edge is a connection between two network nodes. The edge is also called as a link. In networks describing macromolecules the edge is a chemical bond, in protein-protein interaction networks the edge is a physical contact. |
| edge weight | The edge weight represents the strength of the connection between the two nodes. Edge weight is determined by the interaction affinity and probability. |
| network diameter | The diameter of the network is the average of the shortest paths lengths between any two network nodes. |
| scale-free degree distribution | A network has a scale-free degree distribution, if the probability to find nodes with certain number of neighbors (i.e., degree) follows a power-law. Numerically $P = cM^{-\alpha}$, where P is the probability, c a constant, M is the measure, and $\alpha$ is a scaling exponent. Scale-free degree distributions can be best visualized if we take the logarithm of the above equation showing a linear relationship. Nodes with exceedingly large number of neighbors (hubs) have a non-zero probability in scale-free distributions. |
| hub | A hub is a highly connected node of the network, which has a much higher number of neighbors than average. |
| network module | A network module is set of network nodes, which are connected more densely with each other than with their neighborhood (i.e., with other network modules). |
| network core and periphery | If a network has a densely connected substructure (e.g., has a rich club, which is the set of interconnected hubs), we call this densely connected group of nodes as the network core. In these networks non-core nodes belong to the network periphery. |

Most biological networks are small worlds [8] meaning their nodes are very well connected to each other. This is the proverbial "six-degrees-of-separation" (meaning that everyone is on average seven steps away from anyone else in the social network of the entire Earth). In fact, recent data on Facebook showed that this subset of Earth-inhabitants has only 3.74 degrees of separation as an average [9]. If approaching the concept of small worlds in a more general term, small worldness means that the network diameter (i.e., the average number of steps needed to reach a node from another) grows only with the logarithm of the number of nodes and not linearly. Most real world networks have a scale-free degree distribution meaning that their connection structure is uneven allowing some nodes (hubs) to have an unusually large number of neighbors [10]. Networks often have a modular structure, i.e., groups of network nodes (also called as network communities) have denser intra-group connections than the density of the connections of the group with neighboring groups. Biological networks are often hierarchical meaning that nodes and groups have an uneven position in the hierarchical network, which divides them to a 'central core' and a 'periphery' [11, 12].

While the small world character is necessary for the efficient connection and information transmission of biological networks, both hubs and network modules are needed to screen and filter the information. Why is information filtering so important? A complex system (e.g., a cell) receives and generates an extremely large number of perturbations in each second. If all these changes would reach all nodes with the efficiency given by the small world character of their network, our networks would be continuously overwhelmed by information. Thus, network-based information filtering is a major mechanism of learning and memory formation as we will show in Section 4 of this review. Why are hubs and modules helpful in information filtering? Hubs can transmit only a minority of the incoming information at a given time. Network modules due to their dense intra-modular connections and sparse inter-modular contacts keep the incoming information 'trapped' inside the module [3]. 'Gate-keepers' of modular boundaries usually allow the preferential passage of the information to the next module only in special cases, when the network has already been trained to provide a fast transmission of that particular change by previous experience. Inter-modular nodes and connections thus enable the network to 'learn' the consequence of many consecutive inputs, or a single large impulse encoding the novel information to specific transfer pathways.

It is important to note that for the functional analysis of biological networks, network topology is often not enough, but network dynamics also has to be taken into account. In dynamic network models quantities assigned to nodes and/or edges may vary, and/or the background topology itself may also change, where the latter phenomenon is commonly referred as 'network evolution'. Such dynamism may be effi-ciently captured by the concept pair of network rigidity and flexibility, where rigid networks are those, which either structurally or dynamically may not, or do not change, as opposed to flexible networks, which are able to change and/or do change their structure [13]. Modular structure, core and periphery are especially dynamic network parameters as we will show in Sections 3 and 4 of this review.

## 2. MOLECULAR CHAPERONE NETWORKS

Molecular chaperones – as discussed in other papers of this special issue – play a key role in the maintenance of cellular protein homeostasis. They help the folding of *de novo* synthesized proteins, sequester, refold or help the degradation of misfolded proteins, and often chaperone those events of cellular life, which require conformational and functional changes of participating proteins. Thus, chaperones have a paramount role in developmental processes, during/after stress, in combating against diseases and in aging [14-19]. The role of chaperones may become even more important by the recent proposal on the generality of misfolded protein-induced transmissible diseases [20]. Chaperones almost never work alone. They frequently associate with each other forming large chaperone complexes, and (often transiently) bind to other protein complexes of the cell [21-24]. Chaperones also bind to cellular membranes helping their stabilization and dynamics [25-28].

In the following Sections we will first give a brief summary of the protein structure networks of molecular chaperones, where nodes are the amino acids of individual chaperones, and edges are the bonds linking them to each other. Then, we will overview the subset of protein-protein interaction networks molecular chaperones form by their interactions with each other and with their partner or client proteins. Finally, we will describe the special position of chaperones in protein-protein interaction networks.

### 2.1. Protein Structure Networks of Molecular Chaperones

Protein structure networks (often called as residue interaction networks, or amino acid networks) give the molecular background of all other cellular networks at higher levels of hierarchy, such as protein-protein interaction or signaling networks. Protein structure networks have either the individual atoms or whole amino acid side-chains as their nodes. Network edges of protein structure networks are related to the physical distance in the 3D protein structure between amino acid side-chains. This distance is measured between the αC or the βC atoms of amino acids in most cases. Sometimes centers of weight of the side chains are calculated, and distances are measured between these centers. Edges of unweighted protein structure networks connect amino acids having a distance between each other below a cut-off, which is usually between 0.4 to 0.85 nm. Protein structure networks often have weighted links instead of distance cut-offs, where the edge-weight is inversely proportional with the distance between the two amino acid side chains. Covalent bonds may be included or excluded in protein structure networks [11, 29-38].

The molecular structure of chaperones was a subject of several protein structure network studies. Chaperone molecules pose an exciting system from the network point of view due to their flexibility and conformational changes upon ATP or substrate protein binding. The early study of Keskin *et al.* [39] investigated the archetype of molecular chaperone machineries, the GroEL-GroES chaperonin complex using normal mode analysis, which enabled them to tie intra-molecular motions to protein structure network segments. Their method suggested the alternating compression and expansion of the opposing two cavities of chaperonin complexes. ATP binding stabilized a relatively open conformation of the cavities [39]. Later studies on the GroEL-GroES system used Markov propagation (a random process, whose future probabilities are determined by its most recent values) of information transfer through the protein structure network of the molecule, elastic network models and/or normal mode analysis. These studies revealed pathways of intra-molecular propagation of allosteric changes forming a network spanning the whole molecule and involving most ATP and substrate protein binding residues. Inter-modular nodes, hinges, loops and hubs were particularly important in information transmission [40-44].

The Hsp70 machinery also gained a significant attention in protein structure network studies. Comparative studies of the ATP-ase domain with 4 different nucleotide exchange domains using evolutionary and elastic network model methods revealed a set of highly conservative residues involved in nucleotide binding, which participate via a global hinge-bending type of motion in the opening of the ATP-ase domain irrespective of the nature of the bound nucleotide exchanger protein. Moreover, a set of non-conserved, but co-evolved, highly mobile residues was found, which were specific to the nucleotide exchanger. A subset of central residues was also identified near the nucleotide binding site, at the interface between the two lobes of the nucleotide binding domain forming a communication pathway invariant to structural changes [45, 46].

Dixit and Verkhivker [47] combined molecular dynamics simulations, principal component analysis, the energy landscape model and structure-functional analysis of Hsp90 regulatory interactions to systematically investigate functional dynamics of this molecular chaperone. They found a network of conserved regions, which may be involved in coordinating intra-protein allosteric signaling of Hsp90. These motifs may act as cooperating central regulators of Hsp90 inter-domain communications and control of chaperones action. An interactive network of rigid and flexible protein segments was proposed to play a key role in allosteric signaling.

### 2.2. Chaperone Networks

Chaperone networks are subnetworks of protein-protein interaction networks (interactomes) containing molecular chaperones of the given cell type (species) and their physical interactions. Sometimes chaperone networks are extended and also contain the first neighbors of molecular chaperones and their interactions. Chaperone networks are increasingly called as chaperomes.

Protein-protein interaction networks have proteins as their nodes and their direct, physical interactions as edges. These networks are probability-type networks meaning that the interaction strength reflects the probability of the actual interaction. Protein-protein interaction networks may be specialized to the species, to its cell types, to the sub-compartments of the cells, or to certain temporal segments of cellular life, such as a part of the cell cycle, cell differentiation, malignant transformation, etc. These specializations may be direct where the interactions of proteins are experimentally measured in the given species, cell type, cellular compartment, or condition. In many cases the specializations are indirect, where the presence of the actual proteins and/or the intensity of the protein-protein interactions is estimated from the mRNA expression level of the given protein [4-6].

Molecular chaperones form very dynamic complexes with each other, with their co-chaperones and with their substrate proteins. In yeast, a given protein can interact with up to 25 different chaperones during its lifetime in the cell. Chaperones are often forming homo- or hetero-oligomers. Co-chaperones regulate and modify their function by enhancing or inhibiting binding or release of substrate proteins, ATP or ADP [21-24, 48-52].

In yeast cells, two interrelated, but separated chaperone networks have been reported. One of them is called as the CLIPS chaperones meaning the set of chaperones operating to help the folding of *de-novo* synthesized proteins. This subnetwork includes the SSB Hsp70 proteins and the TriC/CCT Hsp60 complex. The other chaperone subnetwork was termed as the HSP chaperone group. Chaperones of this subset mainly assist in the refolding of damaged proteins after stress. This subnetwork contains the SSA Hsp70 chaperone and Hsp90. It is a remarkable self-regulation of yeast cells, that the synthesis of CLIPS chaperones becomes repressed during stress, since cellular proliferation and protein synthesis are inhibited under these conditions to spare energy. On the contrary, synthesis of HSP chaperones is grossly activated after stress [53]. Very interestingly, these major changes were recovered, when the protein-protein interaction network of stressed yeast cells was analyzed [54]. However, the two yeast chaperone sub-networks are not entirely distinct. There are a large number of overlapping chaperones participating in both folding assistance to newly synthesized proteins and misfolded proteins after stress. Examples of such 'dual-mode' yeast chaperones include the SSE1 (Hsp104) chaperone, which acts as a nucleotide exchanger for both key Hsp70 proteins in the different groups [55]. Additionally, members of the yeast Hsp90-related chaperone co-factor complex also have extensive contacts with both Hsp70 complexes [56]. The dual role of Hsp90 was suggested earlier by Young *et al.* [52], who compiled an overlapping network of two chaperone systems for *de novo* protein folding. One of the chaperone systems was called as the "early chaperone network" and contained Hsc70 (the cognate 70 kDa heat shock protein), prefoldin and the Hsp60 complex. Here Hsp90 was an occasional, late component in the folding pathway. The other chaperone system was termed as the "late chaperone network" and contained mostly Hsc70 and Hsp90.

A recent report elaborated further the structure of the yeast chaperone network involving 64 chaperones and 2,691 interacting proteins. The network had 10 modules, where different chaperone-subsets were specialized to different proteins. Chaperone-specificity was also correlated with the copy-number of proteins, which can be rationalized by the fact that low copy number proteins have a higher mutation rate, and therefore may require a more intensive help of chaperones [57].

Similar to the findings of two chaperone sub-networks in yeast cells, in *Escherichia coli* fluorescence resonance energy transfer experiments revealed two chaperone subnetworks specialized to the folding of nascent proteins and the refolding of damaged protein structures after stress. The *de novo* folding sub-network was dominated by Trigger factor, the GroE (Hsp60) machinery and bacterial Hsp90. The stress-related sub-network involved the Clp and Ibp families. The DnaK (Hsp70) family participated in both sub-networks [58]. A recent study elucidated the effect of the two major chaperone complexes (GroE and DnaK) as well as that of Trigger factor on the solubility of ~800 aggregation-prone *E. coli* cytosolic proteins. Both the GroE and DnaK complexes solubilized hundreds of proteins with weak biases. In contrast, overexpression of Trigger factor alone had only a marginal activity. However, when two or all the three chaperones were co-expressed, a subset of proteins was solubilized, which was not rescued by any chaperone alone [59].

It is of interest to compare the above chaperone network of *Escherichia coli* with that of *Mycoplasma* species. *Mycoplasmas* have a high mutation rate and evolve 50% faster than related organisms allowing them an easy escape from the detection and defense mechanisms of the host organism. A likely consequence of this high mutation rate is an increase in the frequency of misfolded *Mycoplasma* proteins. Indeed, estimates using comparative structural genomics resulted in generally lower protein stability of 11 protein families in *Mycoplasma* compared to other bacteria. However, most *Mycoplasmas* have lost either the gene or the activity of their central chaperone, GroEL. This strongly suggests that protein quality control is mostly mediated by protein degradation in these bacteria [60]. This is even more likely, since the alternative *E. coli* folding pathway detected by Kumar and Sourjik [58] contains the Clp family, which includes several major bacterial proteases.

The first comprehensive map of the yeast Hsp90-related chaperome was assembled by Zhao *et al.* [56] containing 198 putative physical interactions and an additional 451 putative genetic and chemical-genetic interactions. Hsp90 was shown to be involved in a large number of cellular functions (including transport processes and several metabolic processes) via a set of different co-factors. These included proteins involved in chromatin remodeling and epigenetic regulation.

A comprehensive study of yeast Hsp90 networks under normal growth conditions and elevated temperature [61] showed that

1. Hsp90 neighbors contained a higher than expected number of hubs;

2. Hsp90 complexes were rather labile indicating a set of low-affinity interactions characteristic to this class of chaperones, and

3. under normal growth conditions, the Hsp90 network was centered on the secretory pathway and cellular transport processes, while under stress the Hsp90 network was more diverse and structured, and became centered around cell cycle regulation, meiosis and cytokinesis.

A recent study compared yeast Hsp90 chaperome with that of the fungal pathogen, *Candida albicans* [62]. In contrast to *S. cerevisiae*, where Hsp90 interacts with more than 10% of the proteome, only 2 interactions of Hsp90 have been identified in *Candida*. Using chemical genomic approach, the study compiled 226 genetic interactors, which – in contrast to yeast cells – were mostly important for fungal growth under specific conditions. Only a few interactors were found to be important in many growth conditions.

Pavithra *et al.* [63] created a chaperone network of the malarial parasite, *Plasmodium falciparum* by combining experimental interactome data with *in silico* analysis. They used interolog mapping to predict protein-protein interactions for parasite chaperones based on the interactions of corresponding human chaperones. The network predicted chaperone functions related to chromatin remodeling, protein trafficking, and cytoadherence. An Hsp90 interacting parasite protein, Cg4, may be associated with drug resistance. The network analysis gave a rational basis for the antimalarial activity of geldanamycin, a well-known Hsp90 inhibitor.

Human Hsp90 chaperone subnetwork became an important subject of studies due to the key role of specific Hsp90 inhibitors, like geldanamycin and its analogues, in targeting cancer [64]. Hsp90 was shown to enrich its local chaperone network in tumors, and acquire a 100-fold higher affinity for its specific inhibitor, 17-allilamido-geldanamycin, than in normal cells [65]. Importantly, Kang *et al.* [66] described that, as opposed to normal cells, mitochondria of tumor cells contain Hsp90 and its homologue, TRAP-1. These chaperones interact with cyclophilin D, an immunophilin inducing mitochondrial cell death, and protect tumor cells from the destructive mechanism of cyclophilin D. Authors have shown that mitochondrium-selective Hsp90 inhibitors may be a novel area of anticancer drug development.

Recently the subnetwork of human Hsp90 containing 1,150 putative nodes and 8,892 edges has been assembled by mining all major protein-protein interaction databases and constructing homologous human interactions by Echeverría *et al.* [67]. Interestingly the "stress response" and the "protein folding" GO terms were represented as only a minor component in the human Hsp90 subnetwork. On the contrary, "development" emerged as a major functional term describing human Hsp90 functions. "Apoptosis", "DNA-repair", "cell cycle", "cytoskeleton", "immune response" "intracellular transport", "lipid and carbohydrate metabolism", "nerve impulse", "protein degradation" "RNA-splicing", "sexual development", "signaling", "stress response" "transcription" and "translation" emerged as more minor components of human Hsp90-related functions [67]. The involvement of Hsp90 in immune functions is supported by the recent finding of its association with STAT3 [68]. Recently an extensive review summarized the potential human Hsp90 sub-interactome compiling the results of more than 23 studies [69]. Their assessment identified RNA-processing, glucose metabolism, the cytoskeleton and extracellular transport as major human Hsp90 network-related functions.

Action of several major chaperones, such as that of Hsp70 or Hsp90, is largely dependent on their co-chaperones. Several recent studies identified chaperone subnetworks by assessing the co-chaperone interactome. Sahi and Craig [70] made a comprehensive analysis of J proteins (which are co-chaperones of the Hsp70 family) in yeast. J proteins were proven to act both as a modulator of the ATPase activity of yeast Hsp70 proteins, and as their anchors to various cellular subcompartments. A recent study identified the network of yeast p23, a key co-chaperone of Hsp90 [71, 72]. Yeast p23 is involved in a broad range of nuclear functions including ribosome biogenesis and DNA repair, as well as in proper Golgi function.

An illustration of the human chaperome (including human chaperones and their interactions) is shown on Fig. (**1**). The core of the network contains 25 chaperone molecules, while the network periphery has 13 chaperone isoforms, a similar organization to that reported earlier by us for the extended chaperome [23]. Most of Hsp60 and Hsp70 chaperones are (and were) in the core, while most of the small heat shock proteins are (and were) in the periphery. The Hsp90 chaperones are divided between the two. The dissection resembles to the duality of yeast "CLIPS-chaperones" (chaperones linked to protein synthesis) and "HSP-chaperones" (stress-induced chaperones) as described by Albanese *et al.* [53].

### 2.3. Position and Dynamics of Molecular Chaperones in Protein-Protein Interaction Networks

As shown in the previous Section, the position of molecular chaperones in protein-protein interaction networks was the subject of intensified research in recent years. In this Section we focus on the knowledge obtained by these numerous studies, highlighting a few important – sometimes hypothetical – special properties of the organization of molecular chaperones in the protein-protein interaction network of the cell.

The first question that comes to mind is: Why molecular chaperone interaction data are not covered better in general interactome studies and in high-throughput experiments? Why is it so general that these inquiries have to be focused to a specific molecular chaperone, or a smaller or larger set of chaperones? The most important part of the answer lies in the fact that most chaperone interactions are filtered out in many high-throughput studies due to the low-affinity, transient binding of molecular chaperones to their partners. Due to this reason chaperone interactions are often not preferentially contained in general interaction databases. This makes the chaperone-directed interactome studies, besides technically challenging, especially important and useful.

Molecular chaperones are not only molecular machines, which help the folding, refolding, activation or assembly of other proteins in a rather localized fashion. Chaperones also have a number of functions, which can be understood only by considering the emergent properties of cellular networks – and chaperones as special network constituents. As an example for this network-related role, the human Hsp90-related chaperome contains almost exclusively such functional modules, where Hsp90 helps the specific cellular function not only as a partner of a particular protein, but as a member of the network module responsible for the function [67].

What makes chaperone positions special in protein-protein interaction networks?

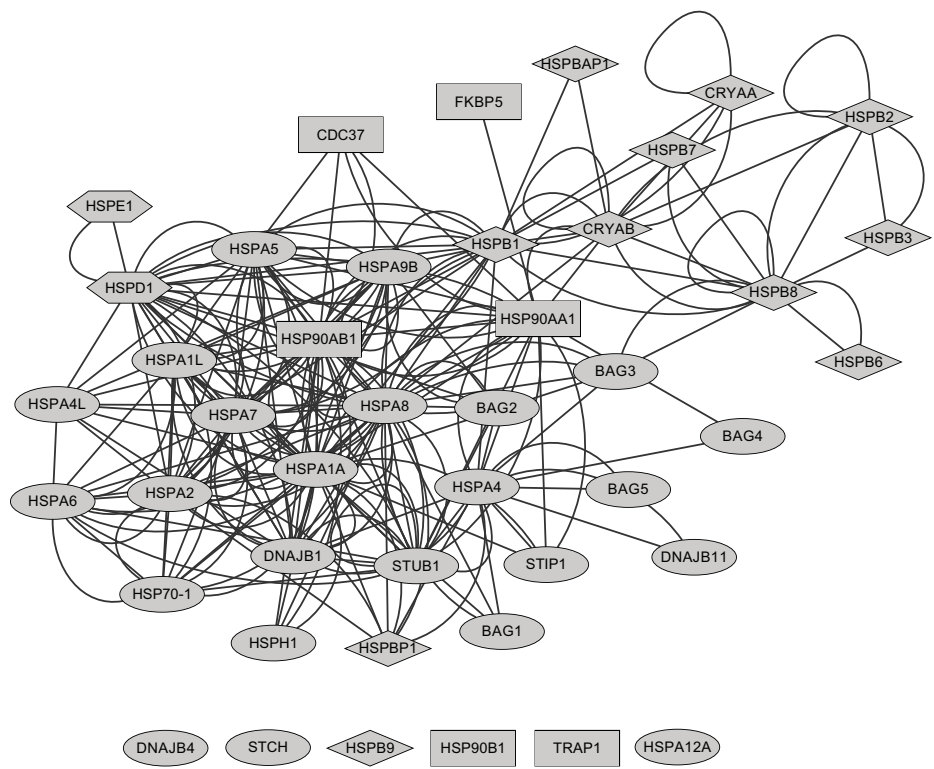

**Fig. (1). The interactome of human chaperones.** Interactions were identified from BioGrid 3.1 ([170], www.thebiogrid.org, version 30.04.2012) and pSTIING ([171], www.pstiing.licr.org, version 30.04.2012). The data set was manually curated. Nodes represent chaperones, while edges are chaperone-chaperone interactions. Node shapes reflect the chaperone classes by molecular weight: small heat shock proteins (diamonds), 60-kDa HSPs (hexagons), 70-kDa HSPs (ellipses), 90-kDa HSPs (rectangles). The network was visualized using Cytoscape 2.8.2 ([172], www.cytoscape.org).

- **Chaperones are usually hubs**, having a much larger number of neighbors than the average.

- Many of chaperone contacts have a low affinity. Due to this low affinity, as well as to the characteristic progress of the ATPase cycle, most chaperone contacts are transient [3, 7, 73]. Therefore, **edges of chaperone contacts have a low weight in protein-protein interaction networks**.

- As a sum of the first two properties: chaperones are not party hubs (using the nomenclature of Han *et al.* [74]) contacting with all their neighbors at the same time. **Chaperones are date hubs** having a continuously changing network position. (Using the nomenclature of Kim *et al.* [75] we may call chaperones as 'singlish-interface' hubs, where many of the binding partners bind to the same binding site one after the other – as opposed to multiple-interface hubs, where a large number of binding sites are anchoring the binding partners in parallel.)

- As it has been shown earlier [74, 76, 77], date hubs quite many times occupy an inter-modular position, meaning that they belong to at least two modules at the same time. **Chaperones are typical inter-modular hubs** in protein-protein interaction networks.

- **Chaperones preferentially connect hubs.** Chaperones are neighbors of several local centers having a lot of second neighbors. This helps the chaperone-mediated cross talk between signaling and gene regulatory pathways enabling chaperones to act as a central switchboard of the cell re-programming cellular functions during and after stress [21-24, 51, 54, 78].

- **Chaperones are in the overlaps of many network modules and occupy a highly dynamic, central position of the interactome.** Chaperones do not only connect two adjacent modules in the interactome (acting like bridges), but also connect many, very distant modules at the same time. Moreover, this connection structure is very flexible. On top of all these, chaperone-mediated changes of far inter-modular contacts encode the integration of the environmental and intrinsic changes observed by the cell. The key multi-modular position gives molecular chaperones a special regulatory role, since they can easily couple, uncouple or even quarantine network communities, i.e., protein complexes, cellular organelles, such as damaged mitochondria, signaling pathways, metabolic routes or genetic regulatory circuits [22, 24, 48, 54, 79].

- **Chaperones are 'creative nodes'.** In 2008 one of the authors (P.C.) called the special position connecting multiple, very distant modules, as the position of 'creative nodes', since it resembles to the position of highly creative people in social networks [80]. Creativity has many forms. As one of them, creative people often have a 'low affinity' to a large number of their neighbors, simply because they get bored by the recurrent ideas they hear in an unchanging environment. If this happens, creative people escape, and explore a new neighborhood. The famous late Hungarian mathematician, Pál Erdős, was an archetype of this personality. Mathematicians who knew Erdős note that only the best mathematicians could maintain Erdős' attention for more than ten minutes. Creativity is not always useful. A creative person may be highly unpredictable and unreliable, since one may never know whether the creative person will accomplish the task he/she was given, or will leave it unfinished, and will jump to another project, which currently excites him/her much more [80, 81]. Chaperones (taken them one-by-one and not as a population) are highly unpredictable and unreliable, too. When the community lives its everyday life in a "business-as-usual" fashion, creative persons are the first who will be fired, due to the unreliability mentioned above. However, the situation will completely change in times of crisis. Here creative nodes behave as the 'life insurance' of the crisis-stuck organization, since they are the only nodes, who have an access to the integrated knowledge of all the organization, and they are the only nodes, who are able to re-combine bits and pieces of these capabilities making up a completely novel solution to survive the crisis and develop afterwards [80, 81]. Chaperones are accomplishing a similar role during and after stress, therefore they participate in network-level adaptation, evolution, and learning and memory formation processes which we will describe in detail in the following Sections.

At the end of this Section on the position and role of chaperones in protein-protein interaction networks, we will highlight a few interesting hot-spots of cellular life, where chaperones play (or may play) an unexpected, novel role. Before doing that, we would like to warn the reader that the list, which follows, is highly subjective, and is, by far, not complete.

- When chaperones position themselves at the intersections of protein-protein interaction networks, they often connect different subcellular compartments (or in other words, organelle networks). As an example, chaperones couple mitochondria to each other and to the endoplasmic reticulum [79, 82]. Several chaperones were shown to associate to membranes [25-28], where they are acting not only as membrane stabilizers, but also connect the cytoplasmic compartment with membrane compartments. The inter-organelle position of chaperones enables them to couple or uncouple nodes of the organelle network. Uncoupling of the organelle network becomes especially important after stress, as it prevents the spread of damage (e.g., in the form of free radicals) from one organelle to the other. Moreover, chaperones may act as an automatic switch to accomplish this task, since during stress, when the amount of misfolded proteins becomes much higher, chaperones become occupied by misfolded proteins, which leads to the dissociation of their original partners [15, 17, 83, 84, M.T. Nguyen and C. Sőti unpublished observations], and causes – among many other consequences – the 'automatic' de-coupling of chaperone mediated inter-organelle contacts. After the first phase of the stress is over, re-coupling may take place again. The behavior of the yeast mitochondrial matrix chaperone, Hsp78, which mediates the restoration of mitochondrial network after stress, is a good example of this latter phase of the process [82].

- Several pieces of evidence on chaperone-mediated regulation of another cellular network, the cytoskeletal network also start to emerge. The yeast chaperone, Hsp104, was shown to be a part of a network linking the actin cytoskeleton with the polarisome and the cytokinesis machinery, and be probably involved in the asymmetric inheritance of damaged proteins by mother and daughter cells [85]. A member of the small heat shock protein family, αB-crystallin was shown to regulate the dynamics of actin filament networks protecting the remaining network integrity after stress [86]. The excellent, recent review of Quintá *et al.* [87] summarized an extensive cytoskeleton-related (tubulin- and actomyosin-related) role of chaperones in cell growth, cell division, cell movement, vesicle transportation, cellular organelle location, localization and distribution of membrane receptors, as well as in cell-cell communication.

- Gong *et al.* [49] identified several cellular hot-spots of molecular chaperone action in yeast cells. They found that yeasts spend more chaperone resources in maintaining the conformational integrity near or close to the nucleus than at other cellular compartments. An enrichment of chaperone-related nuclear functions, such as topoisomerisation, rRNA splicing, ribosome-assembly has also been found. In agreement with these bioinformatic inquiries, Albanése *et al.* [88] found a ribosome-anchored chaperone network of yeast Hsp70 and its two co-chaperones to play an active role in rRNA and ribosome maturation, as well as in the assembly of both ribosomal subunits. Extensive nuclear functions of Hsp90, including chromatin remodeling and epigenetic regulation, have also been described [56, 89, 90]. It is highly possible that chaperones play a key role in the organization of chromatin networks [91].

In summary, chaperones acting as bridges between multiple network modules occupy a highly central position of protein-protein interaction networks. The central network position of molecular chaperones is often extended to higher levels of the hierarchy of cellular networks, such as the membraneous network of cellular organelles, the cytoskeleton and the chromatin network in the nucleus. Moreover, chaperones not only have a central network position, but they are also highly mobile. This enables them to play a key role in the re-organization of protein-protein interaction, signaling, gene transcription, cellular organelle, cytoskeletal and chromatin networks. This chaperone-mediated network re-organization may play a key role in the integration of the stress response, in adaptation, regulation of evolvability, learning and memory formation as we will describe in the next two Sections.

## 3. ADAPTIVE CHANGES OF CELLULAR NETWORKS: THE ROLE OF CHAPERONES

The central position of molecular chaperones in protein-protein interaction networks described in the previous section makes them especially important players in the re-configuration of cellular networks during short- or long-term adaptation, like stress or evolution. These network-level adaptive processes correspond to network-encoded learning and memory formation at the cellular level, and may also play a key role in the molecular mechanisms of learning and memory formation in neuronal networks.

### 3.1. Short-Term Adaptive Changes of Cellular Networks: Changes of Chaperones in Cellular Networks Upon Stress

Many chaperones are also stress, or heat shock proteins, since their synthesis is up-regulated, when the cell experiences stress. During stress, chaperones become increasingly occupied by damaged proteins, and a so-called 'chaperone overload' may easily occur [15, 17]. This 'competitive inhibition' of molecular chaperones may lead to a decoupling the chaperone-mediated inter-modular edges of the protein-protein interaction networks, organelle networks, cytoskeletal network and the chromatin network. De-coupling of network modules may be so extensive that the damaged module becomes quarantined. Since decoupling efficiently prevents the propagation of network damage at the modular boundaries, chaperone-induced module de-coupling provides an additional safety measure for the cell [22, 24, 54, 78, 79].

The above assumptions are in agreement with the results of Luscombe *et al.* [92], who examined the topology of yeast transcriptional signaling sub-networks of 142 transcription factors and 3,420 target genes in five different cellular conditions. The stress-response was governed by a simplified sub-network, which had a shorter diameter and was characterized by large hubs, which probably behaved as integrators of the re-programmed cellular response. On the contrary, the cell-cycle was governed by a highly interwoven, complex structure indicating a multistage internal program [92].

Extending these observations further in a former study [54] we used the ModuLand program package [76] and its Cytoscape plug-in [93] to identify extensively overlapping network modules of the protein-protein interaction network of *S. cerevisiae* under fast-growing conditions and after various stresses. Upon a 15 minute, heat shock at 37°C, the overlap between yeast interactome modules became significantly smaller. In other words the yeast interactome displayed a condensation of its groups, which corresponded to the development of more disjoined large protein complexes upon heat shock and a number of other stress types. In this change the number of inter-modular links decreased [54]. The change of yeast protein-protein interaction networks upon stress showed a stratus → cumulus type of transition [94], where the initial shape of yeast interactome in a resting state corresponded to the flat, dense (dark) low-lying stratus clouds, while the shape of yeast interactome after stress corresponded to puffy (white) altocumulus clouds. More generally, the stratus state resembles a generally flexible state, while the cumulus state has rather rigid intra-modular structure due to the condensation of modules. However, the cumulus state has a rather flexible inter-modular structure due to the partial dissociation of modules. From this point of view the cumulus state has a higher complexity than the stratus state [11, 13, 54].

The partial disassembly of yeast interactome modules is useful, since it

1. spares links and thus energy;
2. slows down the propagation of damage;
3. allows a larger independence of modules and thus a larger exploration radius of their adaptive responses;
4. allows a more adaptive re-organization of the network from pre-formed elements during/upon relief from stress.

Similar modular reorganization is also typical in ecosystems and social networks as the initial phase of their crisis response, and emerged as a general model of adaptive processes [54]. Very importantly, residual bridges between modules of the yeast interactome after stress were maintained by proteins playing a key role in cell survival [54].

When the stress is over, and cellular resources slowly start to get back to normal again, cellular networks may start to re-establish those links, which ceased to operate during stress. Re-gaining of the links shed during stress may slowly reverse the stress-induced stratus → cumulus transition (corresponding to a transition from a generally flexible to a locally rigid, but globally flexible state). This post-stress adaptation phase may slowly re-populate inter-modular edges. Cellular remodeling steps after stress may be greatly helped by the newly synthesized molecular chaperones, since their low affinity interactions effectively sample a large number of proteins, and allow the establishment of a partially novel cellular structure after the stress, as compared to that before the stress. This re-organization of network modules may play a key role in short-term adaptation, learning and memory formation at the level of intra-cellular networks [23, 51, 54, 78].

### 3.2. Long-Term Adaptive Changes of Cellular Networks: The Role of Chaperones in Evolvability

Long-term adaptive processes involve a number of consecutive generations, which are subjects of evolutionary selection processes. Evolvability is defined as the capacity of the system for adaptive evolution. A large evolvability infers a larger potential, a faster speed and a better efficiency to adapt to a new environment [95]. The key paper of Earl and Deem [96] showed that evolvability is a selectable trait, and raised the possibility that evolvability can be regulated by inheritable, encoded mechanisms.

The seminal paper of Rutherford and Lindquist [97] proved the role of Hsp90, a major molecular chaperone in buffering mutation penetrance in fruit flies 15 years ago. Chaperone-induced genetic buffering is released upon stress, which causes the sudden appearance of the phenotype of previously hidden mutations, helps population survival and gives a possible molecular mechanism for fast evolutionary changes. On the other hand, stress-induced appearance of genetic variation at the level of the phenotype cleanses the genome of the population by allowing the exposure and gradual disappearance of disadvantageous mutations by natural selection. Hsp90-buffered changes are remarkably isolated, and can be selected very efficiently and very independently of the expected negative side-effects. Hsp90-buffered traits revealed dozens of normally silent polymorphisms embedded in cell cycle, differentiation and growth control mechanisms [98-100]. Later, these findings were generalized to a large number of other chaperones and other organisms [101-106].

The mechanism by which molecular chaperones may induce and then release genetic buffering seems to be rather complex. The most straightforward explanation of chaperone-induced genetic buffering involves the role of chaperones in protein folding posing that chaperones rescue mutated, misfolded proteins and this is how they buffer the phenotype caused by the compromised action of these mutants. The generality of this explanation was questioned by Bobula *et al.* [107], who first carried out a genome-wide mutagenesis, which was followed by a screen for mutations that were synthetically harmful, when the Hsp70 chaperone system was inactive. Neither the genes identified, nor the nature of genetic lesions implied that the synthetically harmful mutant proteins were chaperone substrates. Later the involvement of Hsp90 in epigenetic and chromatin structure-mediated silencing of existing genetic variants was shown as a potential key factor in mediating Hsp90-induced genetic buffering [108-110]. Importantly, a number of recent publications revealed that a compromise in Hsp90 function inhibit the Piwi-interacting RNA-dependent silencing mechanism leading to transposon activation, and the induction of morphological mutants. (The Piwi-interacting RNA, or in other name: piRNA, is a class of germ-line specific, small RNAs.) At the same time Hsp90 was shown not to affect the short interfering RNA or micro RNA pathways [111, 112]. Sgrò *et al.* [113] showed that the central, charged linker region of Hsp90 acts like a switch of Hsp90-induced genetic buffering in *Drosophila melanogaster*. If the charged linker region became compromised in naturally occurring Hsp90-mutants, the buffering capacity decreased, and the genetic variation was released in a temperature-dependent manner. Tomala and Korona [114] warned that chaperones are involved both in rescuing misfolded mutant proteins by helping them to refold and by eliminating misfolded mutant proteins directing them to degradation. They raised the idea that chaperones are not only buffering genetic changes (by, e.g., re-folding), but may also promote genetic changes (e.g., by directing mutant proteins to degradation, and depriving the cell of the residual activity of these mutant proteins). In their reasoning it is currently not known which part of this balance is stronger in different conditions. Further experimental examples include the chaperone-directed degradation of the mutant CFTR in cystic fibrosis and the mutated glucocerebrosidase in Gaucher disease, where Hsp70 or Hsp90 inhibition enhanced stability and membrane trafficking [115, 116].

In recent years the scientific community has became increasingly aware that not only chaperones but also a large number of other proteins might regulate the diversity of the phenotype. These phenotype regulators may constitute more than 5% of the genome in yeast. It became clearer that a major segment of the modulation of evolvability is encoded by the integrative changes of cellular networks rather than by single bi-molecular interactions. In this context, the central network position of molecular chaperones, and their extremely large dynamics may play a significant role in their influence on evolvability. The remodeling of the inter-modular contacts is an especially intriguing idea for the explanation of the chaperone-mediated sudden changes in the emergent properties of cellular networks (such as in the phenotype of the hosting organism). Different assembly of slightly changed cellular modules may cause profound and abrupt changes of the organisms' functional repertoire — without major gross changes of the underlying structure of protein-protein interactions [3, 7, 51, 98, 99, 117-123].

A recent report [124] showed that the overexpression of heat shock factor-1 (HSF1, the major transcription factor inducing Hsp90 and other major chaperones) leads to an increased stress-resistance in *C. elegans*, which is accompanied by a decreased reproductive fitness and low evolvability. In contrast, in HSF1-deficient mutant worms lower stress-resistance was accompanied by an increased reproductive fitness and by a high evolvability [124]. These opposing trends in stress resistance *versus* reproductive fitness highly resemble to the duality of stratus and cumulus states described in the previous Section. The extreme of the flexible stratus state was described as a "large phenotype" or "s-strategy" having a high proliferation rate and very inefficient energy utilization. Such a system has a low stress-resistance. The locally rigid cumulus state was described as a "small phenotype" or "K-strategy" having a low proliferation rate and highly efficient energy utilization. Such a system has a high stress resistance. The balance between the two types of networks encoding these two segregated phenotypic behavior might be influenced (among others) by the amount of available molecular chaperones [3, 11, 125, 126]. Importantly, neither an extreme form of the flexible stratus state, nor the extreme form of the rigid cumulus state are optimal — as it was shown by Draghi *et al.* [121], who demonstrated that neither absolutely robust (flexible, extremely stratus-type), nor absolutely non-robust (rigid, extremely cumulus-type) systems are successful in attaining a fast adaptation rate. Only balanced systems having both flexibility and rigidity may be successful in long-term adaptation involving both learning and memory formation processes [3, 11, 125, 126] as we will describe in Section 4.B. in more details.

We have only a very few clues how to identify and predict the network positions, which may be occupied by chaperones and/or other modulators of evolvability. C. Ronald Kahn proposed the existence of critical nodes in signal transduction [127]. One of the authors (P.C.) proposed earlier that creative nodes, connecting several distant modules at the same time and having a highly independent and unpredictable behavior, may play a crucial role in this process [80]. Recent advances in network-related methodology identified several novel, complex centrality measures based on non-local network topology (see [77] and references therein). Recently network centrality measures based on perturbation propagation or influencing system-level cooperation have also been introduced [128]. As another recent development, driver nodes were defined making directed networks controllable [129], and other approaches finding key edges and nodes of network control also became available [130, 131]. These novel network measures provide promising approaches to identify the position of key modulators of evolvability in protein-protein interaction and other cellular networks, such as in signaling, gene transcription, cellular organelle, cytoskeletal and chromatin networks.

## 4. NETWORK MECHANISMS OF ADAPTATION, LEARNING AND MEMORY FORMATION

In the preceding Sections we have described the position, dynamics and role of molecular chaperones in cellular networks and the modification of the phenotype (emergent network properties) these networks encode. In this section we will extend the scope of the adaptive changes from the intra-cellular level to the multi-cellular organism level. As a link between the intra-cellular and multi-cellular levels, the adaptive behavior of unicellular organisms was described as an example of learning and memory formation. In one approach, the signal transduction network of bacteria was considered as a learning network of artificial neurons [132]. As we described in Chapter 3.A., the modular reorganization of yeast cells upon various types of stresses provides a general adaptation mechanism in crisis events, which can be regarded as the first step of the learning and memory formation process [54]. In Section 4.A. we will summarize the role of chaperones in learning and memory formation. We will conclude the Section with a – partly hypothetical – description of the role of network topology and dynamics in learning and memory formation.

### 4.1. Role of Chaperones in Learning and Memory Formation

In this section we will summarize the currently available evidence showing the involvement of molecular chaperones in neural processes leading to learning and memory formation. We list the chaperones known to be involved in learning and memory formation in Table **2**.

Chaperones protect the cells against misfolded and aggregating proteins. Neurodegenerative diseases are often characterized by pathological protein aggregation and learning and/or memory deficit. Alzheimer's disease is a classic example, in which amyloid-beta peptide oligomers are thought to be a major component of neurotoxicity. Surprisingly, Alzheimer's disease model knock-out mice defective in the small molecular chaperones, CRYAB and HSPB2, showed better results in context-dependent associative learning, but had locomotion and sensory defects [133]. Alpha-

**Table 2.** Chaperones involved in learning and memory formation processes.

| Chaperone | Localization | Effect on learning and/or memory | References |
|---|---|---|---|
| CRYAB (crystalline) HSPB2, small molecular chaperones | N.D. | Enhanced associative learning | [133] |
| Hsp70 | Mouse hippocampus | Impairment of learning in high-performing and Hsp70-overexpressing mice, enhancement in poor-performing mice | [137, 138] |
| Hsc70 | Rat hippocampus, amygdala, piriform cortex | Protein synthesis during spatial learning and memory | [139] |
| BiP/Grp78 | Mouse hippocampus and cortex | N.D. (complex formation with σ1 chaperone) | [140] |
| σ1 protein | Mouse hippocampus and cortex | Prolonged Ca2+ signaling from ER | [140] |
| A novel DnaJ homolog | Mouse hippocampus | N.D. (a co-chaperone of Hsp70 chaperones) | [169] |
| Trans-membrane glutamate receptor regulatory proteins (TARPs) | *C. elegans* neuron surface Mouse brain extract | Modification of gating current in AMPA-type glutamate receptors | [142] [170] |
| Hsp90 | Organotypic hippocampal slice culture | Trafficking of AMPA-type glutamate receptors into synapses | [143] |
| N-ethylmaleimide-sensitive fusion protein (NSF) | Rat hippocampus | Regulates synaptic transmission of AMPA-type glutamate receptors | [134, 141, 145-149] |
| Sortilin | N.D. | Synaptic plasticity through the regulation of brain-derived neurotropic factor (BDNF) | [150] |
| Heterogenous nuclear ribonucleoprotein (hnRNP) K BAG5 | Rat hippocampus | Improved learning in regular exercise | [168] |

N.D.: Not determined yet.

synuclein is a major constituent of Lewy bodies in Parkinson's disease and interacts with a large number of chaperones. The physiological function of the protein is unclear yet, but *in vitro* experiments suggest that it may be able to prevent the aggregation of various proteins caused by chemical stress [134]. Although this function resembles that of chaperones, currently alpha-synuclein is considered an *in vitro* chaperone only.

Chaperones are involved in the molecular background of caloric restriction. Caloric restriction was shown to affect synaptic plasticity having a potential effect even on learning and memory formation [135].

Hsp70 and Hsc70 were also found in the postsynaptic density [136]. Both Hsp70 and Hsc70 levels increase during learning in the hippocampus [137-139]. The increased Hsp70 chaperone levels may be required for the increased protein synthesis of hippocampal neurons activated by the learning process. Over-expression of Hsp70 results in a learning defect, which may be caused by a sequestration of several proteins important for the dynamical reorganization of the interactome [138].

The sigma-1 receptor regulates the endoplasmic reticulum/mitochondrial inter-organellar $Ca^{2+}$ mobilization through the inositol-1,4,5-trisphosphate receptor. The sigma-1 receptor forms a complex with the major chaperone of the endoplasmic reticulum, BiP (Grp78). Their dissociation leads to prolonged neuronal calcium signaling. Specific agonists of the sigma-1 protein were proven to be effective both in learning- and memory-related processes and in neuroprotection [140].

The AMPA-type ionotropic glutamate receptors are responsible for most of the excitatory synaptic transmission in the brain, and are commonly related to learning. Targeting and delivery of AMPA-type glutamate receptors into synapses affects the synaptic function, maturation and plasticity [141]. Trans-membrane AMPA receptor regulatory proteins are chaperones and obligatory subunits of AMPA-type glutamate receptors and possibly regulate their desensitization [142]. Hsp90 is required for the continuous cycling of AMPA-type glutamate receptors into and from the postsynaptic membrane [143]. Hsp90 may be involved in the formation and disassembly of protein complexes required for AMPA-type glutamate receptor cycling. As we described in Section 2.C., Hsp90 binds to the cytoskeleton, thus it may act like an inter-modular switch between cytoskeletal and AMPA-type glutamate receptor protein communities. Hsp90 was shown to be involved in the acute stress-induced glutamatergic signaling [144]. The N-ethylmaleimide-sensitive fusion protein (NSF) also interacts with AMPA-type glutamate receptors [134, 141, 145-149] sharing the glutamate receptor pool with Hsp90 [143].

Sortilin is an intracellular chaperone, which is involved in the regulation of brain-derived neurotrophic factor (BDNF). It affects the maturation and intracellular localization of brain-derived neurotrophic factor, and it directs its secretory trafficking. The pro-form of brain-derived neurotrophic factor induces neuronal apoptosis and long-term depression. The mature form of the factor regulates neuronal differentiation and long-term potentiation, and has proven to be involved in psychiatric diseases, such as dementia or depression. Sortilin is therefore a hub of the molecular network of both developing and mature neurons [150], and its effect through brain-derived neurotropic factor is manifested in symptoms of a wide range of mental conditions, such as dementia, Alzheimer's disease, depression or obsessive-compulsive disorders.

### 4.2. Network-Related Mechanisms of Learning and Memory Formation

Agnati *et al.* [151, 152] proposed the existence of a global molecular network of brain organization a long time ago. They envisioned a continuum of intra-cellular and extra-cellular molecular networks, which are communicating with each other at special regions of the plasma membrane containing a cluster of receptors involved in learning and memory formation.

The AMPA-type ionotropic glutamate receptors were described in the previous Section as major determinants of learning and memory formation. Recent studies uncovered a growing network of protein-protein contacts of these receptors including the trans-membrane AMPA receptor regulatory proteins, the N-ethylmaleimide-sensitive fusion protein and cytoskeleton associated proteins mentioned in the previous Section, as well as PDZ interacting domain proteins, the neuronal activity regulated pentraxin, Cbln1, the synapse differentiation-induced gene 1 and the cystine-knot receptor modulating protein [153]. Recently, several groups in different organisms revealed the role of actin cytoskeleton modulation via the actin capping adducin protein family in neuronal plasticity [154, 155]. Furthermore, Vukojevic *et al.* [155] presented a link between AMPA-type ionotropic glutamate receptor trafficking and the actin cytoskeleton regulation. The network properties of this growing glutamate receptor-associated interactome, and their chaperone connections are just about to be unraveled and are subject of exciting ongoing studies.

Expanding the glutamate receptor-associated interactome even further, there are many ongoing efforts to explore the synaptic protein-protein interaction network comprising probably as many as 4,000 proteins [156-158]. The EURO-SPIN project (http://www.eurospin.mpg.de/) was established to help understanding the basis of aberrant synaptic transmission at network level aiming to restore normal synaptic function in disease, that is: developing novel therapies for synaptopathies (neurodegenerative diseases, schizophrenia, autism, depression etc.). The Synsys project (www.synsys.eu) aims at molecular analysis of synapse function and dynamic modeling to discover novel pathways and targets enabling therapies for human brain disease. Han *et al.* [159] explored the role of motifs (small, recurring segments of directed networks) in memory formation. Polemans *et al.* [160] established that 10 out of 14 dyslexia-associated proteins form a network involved in neuronal migration and neurite outgrowth. They also predicted 3 additional dyslexia candidate genes from this network context.

Quite a few studies explored the potential dynamics of molecular associations and networks involved in learning and memory formations. A recurring finding was that these networks display bi-stable behavior acting like a switch-type mechanism. An early representation of this was the model of Matsushita *et al.* [161]. In their work the kinetics of the model enzyme network involving the Ca/calmodulin-dependent protein kinase and its associated phosphoprotein-phosphatases was analyzed. They concluded that the parameter space of the system displayed a dual attractor structure, and, with this switch-type of behavior, it was able to act as a short-term molecular memory. A similar dynamics of plasma membrane receptor states was suggested by Agnati *et al.* [151], who proposed that the receptor distribution in the plasma membranes of neuronal cells are centered on a few attractors in the state space. Sossin [162] described that long-term memory formation may arise form the repeated inputs of short-term memory stimuli involving *de novo* protein synthesis, but may also arise as a consequence of a single, strong stimulus involving morphological changes. Song *et al.* [163] examined the long-term facilitation of sensory-motor neuron synapses, and established a protein kinase A-dependent positive feedback loop providing a bistable switch in protein kinase A activity. They proposed that such bi-modal behavior may be a key mechanism of long-term memory formation. Ogasawara and Kawato [164] described a similar bi-modal switch of protein kinase M zeta. Currently we lack a detailed understanding of:

1. How many of these kinase-dependent or other switch-type mechanisms exist;
2. How are they interrelated in a network;
3. How do they lead to more persistent morphological changes; and
4. How these interwoven mechanisms filter the noise and integrate the signals in long-term memory formation.

A network-wide, module-based extension of the above local, motif-like mechanisms of learning and memory formation may be provided by the discrimination of 'stratus' and 'cumulus' type of network as we described in Section 3.A. Here 'stratus' refers to flexible communities, having a high overlap and resembling to a stratus cloud as shown on the left side of Fig. (**2**). 'Cumulus' refers to highly coherent, locally rigid communities, having a low overlap and resembling of an altocumulus cloud as shown on the right side of Fig. (**2**). Bateson *et al.* [125] discriminated between 'large' and 'small' phenotypes of human metabolism, where the large phenotype resembles to the flexibility of stratus net-works, while the small phenotype resembles the local rigidity of cumulus type networks. Adaptation to the decrease of environmental resources (e.g., food) may require as many as 3 generations [3, 11, 125].

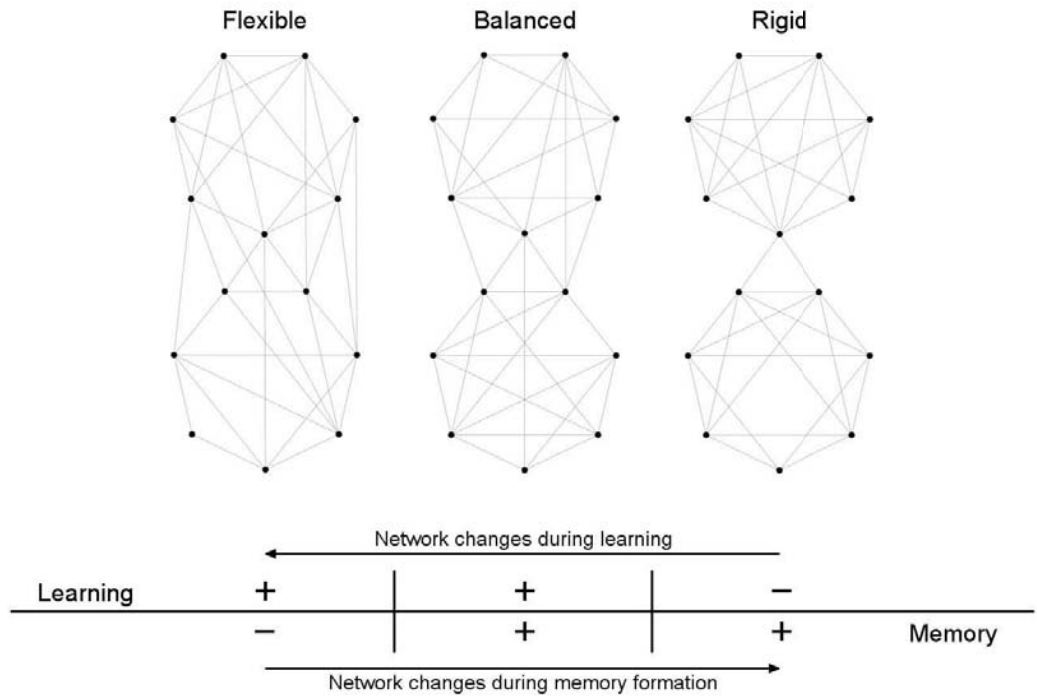

**Fig. (2). Network-related mechanisms of learning and memory formation.** Structure of complex systems often shows a bimodal behavior, where the system is either composed of a highly flexible network (see the illustrative left side of figure, 'Flexible') or has a very cohesive, rather rigid structure (see the illustrative right side of figure, 'Rigid'). Extremely flexible systems may change easily (they 'learn' well). However, these systems cannot preserve the change: they have no 'memory'. On the contrary, extremely rigid systems have difficulties to adapt, to 'learn'. However, once they have changed, they preserve the change (they have 'memory'). The optimal – and most complex – solution is a simultaneous development of network flexibility and rigidity (middle panel, 'Balanced') which is the most successful both to learn and preserve the change, i.e. to adapt to the new situation. In learning networks become more flexible, while during the consolidation of learning, i.e. in memory formation they become more rigid again [165, 166]. Thus the flexible network topology provides 'learning-competent' state. Here networks may have much less modular boundaries than locally rigid, highly modular networks, where the learnt information has already been consolidated in a memory formation process. Since modular boundaries are efficient filters of information, in the 'learning-competent' state information filtering may be much smaller, than after memory formation. This mechanism restricts high information transfer to the 'learning competent' state. After memory formation the stored information is protected by modular boundary-induced segregation and information filtering. Flexible networks of young organisms [126] are generally in a 'learning competent' state. On the contrary, locally rigid networks of old organisms have lost their 'learning competent' state, but store and protect their learnt information efficiently. We note that the above mechanism may operate at the level of both protein-protein interaction and neuronal networks.

To generalize the network-based adaptation, learning and memory formation model we define 'learning' as a capability to encode novel information. We will use the term 'memory' as a capability to store the novel information. Using these terms we propose that extremely flexible systems, as shown on the left side of Fig. (**2**), may adapt easily: they learn well. However, these very flexible systems can not preserve the change: they have no memory. On the contrary, extremely rigid systems, as shown on the right side of Fig. (**2**), have difficulties to adapt, to learn. However, once they have changed, they preserve the change: they have memory. Importantly, neither an extreme form of the flexible stratus state, nor the extreme form of the rigid cumulus state is optimal [3, 11, 13, 121, 125, 126]. To increase their learning ability, extremely rigid systems should increase their flexibility, and shift towards a more balanced system shown the middle of Fig. (**2**). Similarly, to increase their information preservation ability extremely flexible systems should increase their rigidity, and shift towards a more balanced system shown the middle of Fig. (**2**).

We propose that changes of network rigidity and flexibility play a key role in learning and memory formation processes. As an initial evidence of these changes an increased flexibility was shown to occur at the initial phase of cell differentiation, when the cellular networks are in transition in between the original, pluripotent and the final differentiated state [165]. Similarly, a recent study uncovered that network flexibility, defined as the change in the association of a node to different modules of neuronal cells in the human brain, predicts the capability of learning [166].

Importantly, networks in the flexible, 'learning-competent' state may have much less modular boundaries than rigid networks, where the learnt information has already been consolidated in a memory formation process. Since modular boundaries are efficient filters of information, in the 'learning-competent' state information filtering may be much smaller, than after memory formation. Such a mechanism would restrict high information transfer to the state, where it is most needed: to the 'learning competent' state. After memory formation any further incoming information is efficiently filtered, which would provide a very efficient mechanism to protect the stored information.

It is important to note that networks in old organisms are believed to be more rigid than networks of young organisms, which are believed to be more flexible [126]. Taking this notion together with the above hypothesis, and simplifying the complex situation of nature to the extremes, we suggest that flexible networks of young organisms are able to change, thus encode novel information. On the contrary, networks of old organisms are more rigid, thus have more difficulty to encode novel information, but became more efficient to store and protect the information they learnt before.

As we have shown in Section 4.A., chaperones often act directly, as structural components of the molecular machinery involved in learning and memory formation processes. Importantly, synaptic plasticity and long-term potentiation both require large changes in protein levels demanding increased chaperone activity. The exploration of the role of molecular chaperones in the inter-modular changes of large molecular and cellular networks, such as protein-protein interaction networks, signaling, gene transcription, cellular organelle, cytoskeletal and chromatin networks during learning and memory formation will be the subject of further exciting studies.

## CONCLUSIONS AND PERSPECTIVES

In this review we first described the networks of the protein structures of molecular chaperones pointing out the conclusion that amino acids between network modules (communities, domains and sub-domains) of protein structures are key determinants of molecular chaperone action. The importance of inter-modular amino acids in the propagation of conformational changes is a general property of protein structure networks [167].

In Sections 2.B. and 2.C. we described the protein-protein interaction networks of molecular chaperones (also called as chaperomes), and listed the potential properties of chaperone positions in genome-wide interactomes. Chaperones are hubs, have a low intensity and low probability connection structure enabling them to change their neighborhood as the needs of the cell require. Chaperones are occupying an inter-modular position preferentially connecting hubs and other central nodes of several, distant modules. Chaperones behave similarly to highly creative persons in social networks and are therefore good examples of the creative nodes proposed earlier [80]. Chaperones occupy several hotspots of the interactome. They are involved in the reorganization of organelle networks (i.e., networks of mitochondria, the endoplasmic reticulum, the nuclear membrane, the plasma membrane, and other membraneous compartments of the cell) including several intra- and extracellular transport processes. Chaperones participate in the regulation and changes of the cytoskeletal network and link a number of other proteins to both organelle and cytoskeletal networks, as well as connect these two networks with each other. Chaperones have a pivotal role in the nuclear functions, such as in topoisomerisation, rRNA splicing, ribosome-assembly, chromatin remodeling and epigenetic regulation.

In Section 3. we summarized the changes of networks and the role of chaperones in short term adaptive processes as well as in long-term adaptation leading to evolutionary changes. Chaperones (together with many other proteins) play a key role in the reorganization of inter-modular contacts, which is emerging as a central mechanism of the adaptive processes of cells and organisms. We highlighted stratus → cumulus transition as an example of the rearrangement of inter-modular connections. Here 'stratus' refers flexible network communities, having a high overlap and resembling to a stratus cloud as shown on the left side of Fig. (**2**). 'Cumulus' refers to highly coherent, locally rigid network communities, having a low overlap and resembling of an altocumulus cloud as shown on the right side of Fig. (**2**). In our earlier work a stratus → cumulus transition of the yeast interactome was detected after stress [54]. Chaperones have a well-documented role as buffers of a large variety of genetic changes including the occurrence of mutation effects in the phenotype, or transposon activation (see Section 3.B. for more details). It tempting to speculate, that chaperones – mostly *via* their emerging network effects – may actually regulate the level of evolvability of the organism in an integrative manner acting at the system-level.

Finally, we summarized the role of chaperones in learning and memory formation (see Table **2**), and raised the possibility of a number of putative network mechanisms of these processes. We proposed that changes of network rigidity play a key role in learning and memory formation. As we illustrated on Fig. (**2**), flexible network topology provides a 'learning-competent' state. Here network communities are much more linked together than in the locally rigid, highly modular networks. In these latter networks, the emergence of inter-modular boundaries may consolidate the information learnt before in memory formation. In the 'learning-competent' state information filtering may be much smaller, than after the information has safely stored behind emerging modular boundaries. Thus, high intensity, unrestricted information transfer appears there, where it is needed most: in the 'learning competent' state. This state may be characteristic to the flexible networks of young organisms. On the contrary, locally rigid networks of old organisms have lost their 'learning competent' state, but efficiently store and protect the information they learnt before. We note that the above mechanism may operate at the level of both protein-protein interaction and neuronal networks.

Here we highlight the major points, where we predict progress on this rapidly expanding field:

- The advantages of several network representations, such as partially directed networks, colored networks, conditional edges and hypergraphs have not been explored in biological network studies yet;
- The protein structure networks of several chaperones have not yet been assessed;
- The generality of the duality of the *de novo* protein synthesis-related and stress-related chaperone protein-protein interaction sub-networks has not been clarified;
- Human chaperone networks other than that of Hsp90 have yet to be assembled (an experimental verification of the human Hsp90-interactome is missing, too);
- The positions of molecular chaperones in genome-wide protein-protein interaction networks (interactomes) have yet to be assessed;
- The chaperone-related hot-spots of cellular functions have to be examined in multiple systems;
- The changes of protein-protein interaction networks in short-term, and especially in long-term adaptation processes have yet to be clarified;
- The mechanism, by which molecular chaperones may induce and then release genetic buffering is not entirely clear, yet;
- Network-related changes involved in evolvability have yet to be explored;
- The generality of the stratus and cumulus dual network configuration and its transitions should be assessed;

- The properties of the glutamate receptor-associated protein-protein interaction sub-network and its hosting synaptic interactome have yet to be explored;
- The nature and generality of switch-type mechanisms involved in memory formation need more studies; and finally
- The validity of our hypothesis, that changes in network modularity and rigidity and their role in information filtering play a key role in learning and memory formation processes, has to be tested together with its possible connections to the network-based and functional differences between young and aged organisms.

We are at the very beginning of the understanding of short- and long-term (evolutionary) adaptation processes at the systems-level. The network approach offers a great help to understand the integrative mechanisms, how cells and organisms orchestrate these changes, and how molecular chaperones may participate in these processes. The conceptual framework of networks will be an essential tool to understand the growing information on the molecular mechanisms of learning and memory formation processes. We expect a number of exciting discoveries and big surprises in the coming years in this rapidly growing field.

## CONFLICT OF INTEREST

The authors confirm that this article content has no conflicts of interest.

## ACKNOWLEDGEMENTS

Authors would like to thank members of the LINK-group (www.linkgroup.hu) and the Stress-group of the Department of Medical Chemistry at the Semmelweis University, Budapest, Hungary for helpful suggestions. Work in the authors' laboratory was supported by research grants from the Hungarian National Science Foundation (OTKA-K83314), from the EU (TÁMOP-4.2.2/B-10/1-2010-0013) and by a residence at the Rockefeller Foundation Bellagio Center (PC).